\documentclass[aps,prl,nofootinbib,amsmath,twocolumn,preprintnumbers]{revtex4-1}
\usepackage{amssymb,esvect,amsmath,graphicx,latexsym,amsthm,slashed,eso-pic,hyperref}
\usepackage[hang,flushmargin]{footmisc}
\usepackage{tikzfeynman}
\usepackage{placeins}
\usepackage{bbold}
\usepackage{multirow}
\usepackage{slashed}
\usepackage{cancel}

\newcommand{\beq}{\begin{equation}} \newcommand{\eeq}{\end{equation}}
\newcommand{\bea}{\begin{eqnarray}} \newcommand{\eea}{\end{eqnarray}}

\begin{document}
\preprint{YITP-SB-18-28}

\title{Aligned and Spontaneous Flavor Violation}

\author{Daniel Egana-Ugrinovic$^{1}$}
\author{Samuel Homiller$^{1,2}$}
\author{Patrick Meade$^{1}$}
\affiliation{$^{1}$C. N. Yang Institute for Theoretical Physics, Stony Brook University, 
 Stony Brook, NY 11794\\
 $^{2}$Physics Department, Brookhaven National Laboratory, Upton, NY 11973}

\begin{abstract}
We present a systematic spurion setup called Aligned Flavor Violation (AFV) that allows for new physics couplings to quarks that are aligned with the Standard Model (SM) Yukawas, 
but do not necessarily share their hierarchies nor are family universal.
Additionally, 
we show that there is an important subset of AFV called Spontaneous Flavor Violation (SFV),
which naturally arises from UV completions where the quark family number and CP groups are spontaneously broken.
Flavor-changing neutral currents are strongly suppressed  in SFV extensions of the SM.
We study SFV from an effective field theory perspective and demonstrate that SFV new physics with significant and preferential couplings to first or second generation quarks may be close to the TeV scale.

\end{abstract}
\maketitle

\section{Introduction}
Strong constraints on  Flavor-Changing Neutral Currents (FCNCs) and CP violation (CPV) indicate that, naively, 
the scale of new physics needs to be above $\sim 10^5\,\mathrm{TeV}$  \cite{Bona:2007vi}, well beyond the energy frontier (EF) reach of current and future experiments. To reconcile this disparity of scales,  assumptions need to be made for the form of any couplings of beyond the SM (BSM) particles to the quarks of the SM.  For example, the most conservative assumption could be that BSM states always couple to the SM in a flavor-blind manner.  However, this inherently biases the search for BSM physics.  Flavorful BSM assumptions can also be made, but these too represent implicit biases for BSM searches at the EF.

While a complete model of flavor is desirable, the dynamical scales associated with such a model may be $\gg$~TeV scale.  As such, a full model of flavor is not needed to describe the TeV scale, and the required assumptions about flavorful BSM physics can be encoded through EFT and/or spurion methods.  The most common assumption is Minimal Flavor Violation (MFV)~\cite{DAmbrosio:2002vsn},
which imposes that in extensions of the SM, the breaking of the SM flavor group is entirely from SM Yukawa spurions.
By construction, 
MFV inherits much of the flavor protection that exists in the SM, 
e.g., CKM and GIM suppression of FCNCs.
This allows for the scale of new physics to be close to the TeV scale.   
Strictly speaking, MFV applies to the EFT formalism, but the same spurion reasoning can be applied to specific cases of BSM physics.  In turn, MFV makes a precise set of predictions, 
the strongest one being that the couplings of new physics to quarks are either flavor blind, 
as in theories of gauge mediation \cite{Giudice:1998bp},
or retain the hierarchies of the SM Higgs Yukawa couplings. 
Other extensions of MFV such as GMFV \cite{Kagan:2009bn} and NMFV \cite{Agashe:2005hk} retain these properties and primarily affect only third-generation physics. 
Thus, while these extensions have EF implications for third-generation BSM searches,
they generically provide no more guidance than naturalness considerations. 

From the perspective of EF searches, especially at the LHC, it is important to make sure that BSM physics is not missed because of theoretical assumptions.  Therefore to extend the scope of flavorful EF searches, we introduce a new flavorful Ansatz: Aligned Flavor Violation (AFV).  
 AFV allows for phenomenologically interesting couplings to the first and second generation quarks, and/or highly generation dependent couplings,
 with the defining property that all flavor-changing processes are CKM suppressed. 
 While the basic concept of ``alignment" in flavor physics has long been understood in particular models~ \cite{Nir:1993mx,Leurer:1993gy,Branco:1996bq,Penuelas:2017ikk}, we generalize it to the spurion perspective of the SM flavor group.  This allows one to define AFV theories either as an EFT description which can be studied in its own right, or a prescription for how any particular BSM states can couple to quarks as with MFV.

The definition of AFV requires that the background expectation values of new flavored spurions take on a particular form.
From the IR point of view this can be viewed as a systematic attempt to understand aligned Ans\"atze for all BSM theories.  
However, without a UV completion it is not necessarily any more robust or compelling than, e.g., 
picking a Yukawa texture that suppresses FCNCs for instance in an extended Higgs sector ~\cite{Cheng:1987rs}.  
Additionally, CKM suppression alone is generically not sufficient to reduce the bounds from flavor physics to LHC energies.

While AFV is an interesting systematic generalization of flavor alignment, 
there is a class of AFV theories that we refer to as Spontaneous Flavor Violation (SFV), 
which is much more compelling.
In SFV theories the quark family number and CP groups are broken only by wave-function renormalization of the SM right-handed up or down quarks (leading to two types of SFV, up-type or down-type).
From the IR perspective, 
SFV theories restrict the set of allowed AFV spurions,
resulting in both CKM \textit{and} Yukawa suppression of FCNCs, 
as in MFV.   
Nevertheless, 
this is still not any more compelling than selecting an ad-hoc IR texture for the spurions to suppress FCNCs.
However, 
from the UV perspective,
the SFV subset is strongly motivated since it is naturally selected by a simple class of UV completions.
One example of such UV completions are theories where the quark family number and CP groups are broken spontaneously in some extended flavored sector, 
and the breaking is communicated to the SM right-handed quarks via mixing with heavy vector-like quarks.
Because these UV completions lead to CP violation only in wave-function renormalization, 
they are all theories that provide a UV solution to the strong CP problem.
Examples of such theories are Nelson-Barr \cite{Nelson:1983zb,Barr:1984qx,Barr:1984fh,Bento:1991ez} and Hiller-Schmaltz models  \cite{Hiller:2001qg}.

In this letter we demonstrate the basic setup of AFV and SFV theories, 
we derive the generic bounds on SFV theories from an EFT point of view,
and we present the example UV completion indicated above for SFV.
Just as with MFV or AFV,  SFV can be applied to any particular BSM model to determine the couplings to SM quarks.  
In a forthcoming paper, we show that in a 2HDM,
the SFV Ansatz allows for extra Higgs states at the $\sim 100 \, \textrm{GeV}$ scale with Yukawa couplings $\gg y_b^{\textrm{SM}}$ to {\em any} specific quark generation,
while retaining consistency with flavor bounds.  The SFV Ansatz therefore opens up many opportunities for the exploration of new physics at both B-factories and high energy colliders.

\section{Aligned Flavor Violation}
\label{sec:nonuniversal}
To define AFV in as general of possible way, we exploit the flavor and reparametrization symmetries of the SM after EWSB.
The quark flavor symmetry group of the SM, $U(3)_q^3\equiv U(3)_Q \times U(3)_{\bar{u}} \times U(3)_{\bar{d}}$, is broken to $U(1)_Y \times U(1)_B$ by the background values of the Yukawa spurions
\begin{equation}
\mathcal{L} \supset  
-{y}^u_{ij} ~ Q_i H \bar{u}_j
+  {y}^{d\dagger}_{ij} Q_i H^c \bar{d}_j 
\, ,
\end{equation}
which have definite transformation properties under $U(3)_q^3$.
We define the singular value decomposition of the SM Yukawas as
\begin{eqnarray}
\nonumber
y^{u}  & = & 
U_{Q_u}~\! Y^u ~\! U_{\bar{u}}^\dagger
\equiv
U_{Q_u}~\! \textrm{diag}(y_u^{\textrm{SM}},y_c^{\textrm{SM}},y_t^{\textrm{SM}}) ~\! U_{\bar{u}}^\dagger
\quad ,
\\
y^{d\dagger} 
& = & 
U_{Q_d} ~\! Y^d ~\! U_{\bar{d}}^\dagger
\equiv
U_{Q_d} ~\! 
\textrm{diag}(y_d^{\textrm{SM}},y_s^{\textrm{SM}},y_b^{\textrm{SM}}) 
~\! U_{\bar{d}}^\dagger
\quad .
\label{eq:lambdau1}
\end{eqnarray}
The unitary matrices above transform between a generic flavor basis and the quark mass eigenbasis, where in the SM the Yukawa interactions are flavor diagonal. 
The mass eigenbasis is defined only up to a $U(1)_{R}^6=U(1)_{R}^5 \times U(1)_B$ reparametrization group defined by 
\begin{eqnarray}
\nonumber
U_{Q_{u}(\bar{u})} &\rightarrow& U_{Q_{u}(\bar{u})} \, \textrm{diag}(e^{i\alpha_u},e^{i\alpha_c},e^{i\alpha_t}) 
\quad ,
\\
U_{Q_{d}(\bar{d})} &\rightarrow&  U_{Q_{d}(\bar{d})}\, \textrm{diag}(e^{i\alpha_d},e^{i\alpha_s},e^{i\alpha_b})
\quad .
\label{eq:reparametrization}
\end{eqnarray}
The $U(1)_{R}^5$ factor of the reparametrization group is \textit{independent} of the flavor group, 
and can be understood as a symmetry of moving to the mass basis.
Physical observables must be flavor and reparametrization invariant.
The only non-trivial flavor-invariant combination of the unitary matrices in Eq. \eqref{eq:lambdau1} is the CKM matrix,
defined as 
\begin{equation}
V \equiv U_{Q_u}^T U_{Q_d}^* \quad .
\label{eq:VCKM}
\end{equation}
However, the CKM matrix transforms non-trivially under $U(1)_{R}^5$, 
so a choice of reparametrization basis can remove 5 phases leaving the unique physical CKM phase. 

The MFV hypothesis states that only tensor products of $y^u$ and $y^d$ characterize flavor violation.  
To go beyond this, 
we introduce new flavored spurions.
For instance, we may introduce flavor spurions $\kappa^u$ and $\kappa^d$, 
with the same transformation properties under $U(3)_q^3$ as $y^{u,d}$.  
In the SM EFT or in a generic BSM theory, 
this would imply large FCNCs, 
therefore we require additional restrictions on the flavored spurions. 

An inherent feature of the SM at scales beneath EWSB is that the $U(1)_{R}^6$ symmetry ensures that Yukawa interactions are diagonal in the mass eigenbasis and $V$ is the only source of flavor-changing processes.  
CKM suppression of flavor-changing effects can be retained if we impose the condition that the only flavor invariant spurion that transforms non-trivially under $U(1)_{R}^6$ is $V$ itself.
This also guarantees that  the CKM matrix contains only one physical phase.
We define this flavor setup as Aligned Flavor Violation (AFV).  
All AFV spurions may then be expressed as an expansion in powers of the CKM matrix,
which we refer to as the alignment expansion. 
As an illustration, 
the most general AFV spurions $\kappa^{u,d}$ consistent with the flavor and reparametrization symmetries, 
up to second order in the alignment expansion take the form
\begin{eqnarray}
\nonumber
\kappa^u
&=&
U_{Q_u} 
\,
\Big[
~
K^{u}
+
K^{u'}
V^*
K^{u'}
V^T
K^{u'''}
+
\mathcal{O}
\big(\, V^4 \,\big)
\Big]
U_{\bar{u}}^\dagger
\quad ,
\\
\kappa^{d\dagger}
&=&
U_{Q_d}\, 
\Big[
~
K^{d}
+
K^{d'}
V^T 
K^{d''}
V^*
K^{d'''}
+
\mathcal{O}
\big(\, V^4 \,\big)
\Big]
\, 
U_{\bar{d}}^\dagger
\quad ,
\nonumber
\\
\label{eq:kukddefinition}
\end{eqnarray}
where $K^{x}$ are the matrix coefficients of the alignment expansion and are arbitrary flavor invariant, complex-diagonal $3\times 3$ matrices.
They must be diagonal to ensure that they are $U(1)^5_R$ invariant, 
as required by the AFV Ansatz.
$U(1)^5_R$ invariance also makes their phases physical CP-violating phases.
New physics couplings to the three quark generations via the spurions $\kappa^{u,d}$ are neither flavor blind nor do they respect the hierarchies of the SM Yukawas necessarily.
This is in stark contrast to MFV.

AFV spurions in all baryon-number representations may be obtained by taking tensor products of these basic spurions Eqns.  \eqref{eq:kukddefinition},
and they may also be expressed as an expansion in powers of the CKM matrix.
Linear combinations and tensor products of AFV spurions are also aligned.
In any new physics theory, renormalization group evolution (RGE) only renormalizes the alignment expansion matrix coefficients,
so Flavor Alignment is radiatively stable.

While AFV is an interesting generalization of MFV, it has an important shortcoming.
From the UV perspective there is no obvious symmetry reason why AFV should be realized, 
since the $U(1)_R^5$ group is only an auxiliary group redefining quark mass eigenstates.  
In a given model there may be a symmetry one can identify to guarantee CKM suppression of FCNCs, e.g. ~\cite{Nir:1993mx,Leurer:1993gy,Branco:1996bq,Antaramian:1992ya}, 
but as mentioned earlier, 
CKM suppression alone is not sufficient to realize new physics at LHC energies.  
In practice, 
by symmetries or assumption, 
models that realize an AFV Ansatz select
only the lower order terms in the alignment expansion, 
aiming for simultaneous diagonalizability of the flavor spurions with the SM Yukawas, 
examples being ~\cite{Nir:1993mx,Leurer:1993gy,Penuelas:2017ikk} for SUSY or the 2HDM.
Other examples of flavorful theories with scalars coupling to quarks are \cite{BarShalom:2007pw,Giudice:2011ak,Bar-Shalom:2018rjs}.

While it is still interesting to study the implications of AFV, 
the above shortcomings make AFV theories less appealing for general EF searches.  
In the next sections we address these issues by showing that there is a  subset of AFV that we refer to as SFV, 
which has additional suppression beyond CKM {\em and} arises from a simple class of UV completions.

\section{Spontaneous Flavor Violation}
\label{sec:SFV}
A class of AFV extensions of the SM called SFV is realized if at a UV boundary scale $\Lambda_{\textrm{BC}}$,
flavor-changing processes and CP breaking are introduced exclusively via wave-function renormalization of right-handed SM quarks.
While there is a straightforward UV completion of this SFV Ansatz above $\Lambda_{\textrm{BC}}$, 
we first show how defining SFV in this way naturally leads to additional suppression of FCNCs beyond CKM suppression.  
This allows for non-trivial flavored BSM physics at low energies, 
without having to worry about the explicit UV completion that we discuss in a later section.  

To be more explicit, we define an SFV theory by the following two conditions
at the boundary scale $\Lambda_{\textrm{BC}}$:
\begin{enumerate}
\item the only renormalizable interaction in the theory breaking the $U(1)_f^3\times \textrm{CP}$ symmetry is wave-function renormalization of either the right-handed up ($\bar{u}$) \textit{or} down-type ($\bar{d}$) SM quarks \textit{and}
\end{enumerate}
\begin{enumerate}
\setcounter{enumi}{1}
\item 
the theory contains no flavor breaking spurions or fields appearing in renormalizable interactions that transform under $U(3)_{\bar{u}}$ or $U(3)_{\bar{d}}$ correspondingly, 
besides the SM ones and the wave-function matrix above.
\end{enumerate}
We refer to theories satisfying the conditions 1. and 2. as up- or down-type SFV depending on the quark being renormalized,
regardless of the particular UV completion leading to the first condition.
We dedicate the rest of this section to show that in up(down)-type SFV,
all the new flavor spurions are aligned,
all down(up)-type FCNCs are suppressed by factors of SM Yukawas entering in the combinations $V^T Y_{u}^2 V^*$ ($V^* Y_{d}^2 V^T$),
and in addition, new SFV physics may have generic family non-universal couplings to SM quarks.
Imposing that only wave-function renormalization breaks CP is not required for alignment,
but ensures that any UV completion leading to the SFV Ansatz solves the strong-CP problem \cite{Hiller:2001qg}.

We now demonstrate the features above in up-type SFV. 
The proof for down-type SFV is obtained by interchanging the up and down type quarks in what follows.
In up-type SFV, 
at the scale $\Lambda_{\textrm{BC}}$, the Lagrangian contains the interactions
\begin{eqnarray}
\nonumber
\mathcal{L}
&\supset&
i
Z^u_{ij} \bar{u}_i^\dagger
\bar{\sigma}^\mu D_\mu
 \bar{u}_j
+
i
\bar{d}_i^\dagger
\bar{\sigma}^\mu D_\mu \bar{d}_i
+
i
\bar{Q}_i^\dagger
\bar{\sigma}^\mu D_\mu
 \bar{Q}_i
 \\
 &-&
 \big[
{\eta}^u_{ij} ~ Q_i H \bar{u}_j
-  {\eta}^{d}_{ij} Q_i H^c \bar{d}_j 
+
\textrm{h.c.}
\big]
+\mathcal{L}_{\textrm{BSM}}
\quad ,
\label{eq:SFVstructure}
\end{eqnarray}
where without loss of generality we work in the canonical kinetic basis for $\bar{d}$ and $Q$ and we omit other SM interactions without quarks that are not relevant for our discussion.
The term $\mathcal{L}_{\textrm{BSM}}$ represents operators involving any new physics and SM fields,
with the defining limitation that no new spurions or fields transforming under $U(3)_{\bar{u}}$ are allowed at the renormalizable level.
As a consequence, 
all renormalizable interactions involving the up-type right-handed quarks contain the product ${\eta}^u_{ij} \bar{u}_j$.
The mass scale $\Lambda_{\textrm{NP}}$ of the new physics states with renormalizable interactions to the SM may be accessible at colliders.
Such new physics may for instance be part of a TeV scale sector providing a solution to the hierarchy problem. 
Non-renormalizable operators may appear in $\mathcal{L}_{\textrm{BSM}}$
but are suppressed by mass scales related to the SFV UV completion that are similar to $\Lambda_{\textrm{BC}}$ or heavier, 
which are assumed to be inaccessible at current collider and flavor experiments and are neglected in what follows.
\footnote{
One trivial way to justify neglecting non-renormalizable interactions in $\mathcal{L}_{\textrm{BSM}}$,
is to require $\Lambda_{\textrm{BC}} \gtrsim 10^5 \, \textrm{TeV}$. 
However, 
the higher dimensional operators appearing in $\mathcal{L}_{\textrm{BSM}}$ and their Wilson coefficients depend on the specific SFV UV completion, 
so the mass scale of such operators and therefore $\Lambda_{\textrm{BC}}$ may in practice be $\ll 10^5 \, \textrm{TeV}$. }
Finally and also by definition,
in Eq. \eqref{eq:SFVstructure}
there is a special flavor basis that we commit to, 
in which $Z^u_{ij} $ has off-diagonal complex entries
and the matrices $\eta^{u,d}$ are real-diagonal.
Additional allowed spurions in \textit{any} $U(1)_B$ preserving representation of the $U(3)_Q \times U(3)_{\bar{d}}$ group that may exist in $\mathcal{L}_{\textrm{BSM}}$, 
can be written as tensor products of arbitrary real-diagonal spurions transforming as the down-type SM Yukawa.

To go to the fully canonical kinetic basis,
we define the square-root like matrix $\sqrt{Z^u}$ by
\begin{equation}
Z^u=\sqrt{Z^u}^{\dagger} \sqrt{Z^u} \quad \quad ,
\label{eq:sqrtmatrix}
\end{equation}
and redefine the up-type quarks
\begin{equation}
\bar{u}'_i
=
\big(
\sqrt{Z^u}
\,\big)_{ij} 
\, 
\bar{u}_j
\quad .
\label{eq:fieldredef}
\end{equation}
Dropping the primes on the redefined quark fields, 
the renormalized Lagrangian is
\begin{eqnarray}
\nonumber
\mathcal{L}
&
\supset
&
i
\bar{u}_i^\dagger 
\bar{\sigma}^\mu D_\mu
\bar{u}_i
+
i
\bar{d}_i^\dagger
\bar{\sigma}^\mu D_\mu
 \bar{d}_i
+
i
\bar{Q}_i^\dagger 
\bar{\sigma}^\mu D_\mu
\bar{Q}_i
\\
&-&
\big[
{y}^u_{ij} ~ Q_i H \bar{u}_j
-  {y}^{d^\dagger}_{ij} Q_i H^c \bar{d}_j 
+
\textrm{h.c.}
\big]
+
\mathcal{L}_{\textrm{BSM}}
\,\,,
\end{eqnarray}
where the renormalized Yukawas are
\begin{eqnarray}
{y}^u
&=&
 {\eta}^u
 \big(
\sqrt{Z^u}
 \, \big)^{-1}
 =
V^T~\! Y^u ~\!  
\quad ,
\label{eq:flavorbasisup}
\\
{y}^{d\dagger}
&=&
 {\eta}^d
 =
Y^d
\quad .
\label{eq:flavorbasisdown}
\end{eqnarray}
The basis Eqns. \eqref{eq:flavorbasisup},\eqref{eq:flavorbasisdown} corresponds to the flavor basis obtained by setting $U_{\bar{u},\bar{d},Q_d}=\mathbb{1}, U_{Q_u}=V^T$ in Eq. \eqref{eq:lambdau1}.
Note that the off-diagonal terms of the CKM matrix are exclusively due to wave-function renormalization.
The SM down-type Yukawa matrix and all the rest of the spurions appearing in $\mathcal{L}_{\textrm{BSM}}$ transforming only under $U(3)_Q \times U(3)_{\bar{d}}$ are not renormalized,
while all SM and BSM interactions involving the up-type right-handed quarks end up containing the product $y^u_{ij} \bar{u}_j$.
As a consequence, 
in our flavor basis the allowed spurions are the up-type SM Yukawa Eq. \eqref{eq:flavorbasisup}, 
the real-diagonal down-type SM Yukawa Eq. \eqref{eq:flavorbasisdown}
and spurions appearing in $\mathcal{L}_{\textrm{BSM}}$,
which can be expressed as tensor products of arbitrary real-diagonal down-type Yukawas.
We conclude that the theory is flavor aligned.
In particular, 
SFV only selects the lowest order terms in the alignment expansion, c.f. Eq. \eqref{eq:kukddefinition}.
In addition, 
since all the down-type quark FCNCs are formed with the bilinears $Q_i^\dagger Q_j$, $Q_i \bar{d}_j$ and $\bar{d}_i^\dagger \bar{d}_j$,
and since insertions of $y^u y^{u\dagger}$ are required in these bilinears to obtain down-type FCNCs, 
all down-type FCNCs are suppressed by factors of $(V^T Y_{u}^2 V^*)_{ij}$.

In summary,  in the up-type SFV Ansatz all new spurions transforming under $U(3)_Q \times U(3)_{\bar{d}}$ in our  flavor basis Eqns. \eqref{eq:flavorbasisup},\eqref{eq:flavorbasisdown} are arbitrary $3\times 3$ real-diagonal matrices or tensor products of such matrices. 
No new spurions transforming under $U(3)_{\bar{u}}$ are allowed, 
so new physics couplings to the right-handed up-type quarks are either flavor blind or formed out of the SM up-type Yukawa.
The down-type SFV prescription is similar,
but with up and down-type right-handed quarks interchanged. 
The CKM phase is the only CP violating phase in SFV.

Note that SFV is not stable under RGE below the scale $\Lambda_{\textrm{BC}}$, 
but RGE corrections are strongly suppressed by both loop and CKM factors in any new physics model.
For a stability analysis in similar theories see \cite{Gori:2017qwg,Penuelas:2017ikk,Botella:2018gzy}.
For a study of the stability of UV solutions to the strong-CP problem see \cite{Dine:2015jga}.

\section{Flavor bounds in an example SFV theory}
\label{sec:flavor}
As a simple application of the SFV Ansatz, 
consider extending the SM flavor spurion content with only one new down-type Yukawa spurion $\kappa^d$.
In up-type SFV, 
$\kappa^d$ is guaranteed to be flavor aligned,
and in the flavor basis used in the previous section it is also real-diagonal,
\begin{eqnarray}
\kappa^{d\dagger} & = & K^d  \equiv \textrm{diag}(\kappa_d,\kappa_s,\kappa_b)
\quad , \quad \kappa_f \in \mathbb{R}
\quad .
\label{eq:sfvcondition}
\end{eqnarray}
where $\kappa_{d,s,b}$ are arbitrary Yukawa couplings.
Such a spurion may for instance couple a second Higgs doublet  \cite{Davidson:2005cw}, heavy vector-like quarks  \cite{Grinstein:2010ve} or new gauge bosons \cite{Langacker:2000ju,Grinstein:2010ve,Fox:2011qd,Carone:2013uh} to SM quarks.

To assess how effective the SFV Ansatz is in suppressing FCNCs for a generic new physics theory,
we make use of an EFT approach and explore constraints on dimension six operators.  In such an EFT the
Wilson coefficients are controlled by products of SM Yukawas and  $\kappa^d$,
as illustrated in Table  \ref{tab:operatorbound}.
From Table  \ref{tab:operatorbound} we see that all FCNCs are suppressed by CKM factors,
as expected from any flavor aligned theory,
and
all down-type FCNCs come with factors of $(V^T Y_{u}^2 V^*)$ as previously stated. 

In table \ref{tab:operatorbound2} we present the experimental bounds on the scale of the dimension six SFV operators.
For comparison, 
we also show bounds on dimension-six MFV and flavor anarchic operators.
Since MFV operators are also allowed by definition in an SFV theory, 
values of $\kappa_{d,s,b}$ leading to $\Lambda^{\textrm{SFV}}_{\textrm{NP}} \leq \Lambda^{\textrm{MFV}}_{\textrm{NP}}$ requires us to
take the MFV limits instead. 
Bounds on SFV operators are much weaker than on generic flavor-anarchic new physics.
Importantly, 
the scale at which new SFV physics may be found consistent with flavor bounds is \textit{generation specific},
since it depends on the three new Yukawas $\kappa_{d,s,b}$ independently.
As an example, 
consider a scenario in which new physics is mostly coupled to first generation quarks.
For concreteness, take $\kappa_{s,b}=0$, and $\kappa_d \sim 10^5 \, y_d^{\textrm{SM}} (\sim 0.1)$.
From Table \ref{tab:operatorbound2} we see that new physics with such non-universal couplings to first generation quarks may be close to the TeV scale. 
\begin{table} [ht!]
\begin{center}
$
\begin{array}{c|c}
   \text{Operator}
  &
  \text{SFV factor}
    \\ \hline\,
(Q_1^\dagger \bar{\sigma}^\mu Q_2)^2
&
\begin{array}{c}
C_D^1=(\, V^* K_d^2 V^T \,)^2_{12} 
\\
C_K^1
=
(\, V^T Y_u^2 V^* \,)^2_{12} 
\end{array}
\\
\multirow{2}{*}{
$
(Q_1 \bar{d}_{2})(Q_2^\dagger \, \bar{d}_1^\dagger)
$}
&
\Big[ (\, V^T Y_u^2 V^* K^d)_{12}
\\
&
(\, V^T Y_u^2 V^* K^d)^*_{21}
\Big]
\\
Q_2
H^c
 \,
 \sigma^{\mu \nu}
  \bar{d}_3 
 \,
 F_{\mu \nu} 
&
\big[
\, (V^T Y_{u}^2 V^*) \,
 K^d
 \,
 \big]_{23}
\\
\end{array}
$
\end{center}
\caption{
Selection of dimension-six FCNC operators with their SFV coefficients.}
\label{tab:operatorbound}
\end{table}

\label{sec:UVcompletion}

\begin{table} [ht!]
\begin{center}
$
\begin{array}{c|ccc}
   \mathrm{Operator}
    &  
      \Lambda_{\text{NP}}^{\textrm{anarchic}} \, \mathrm{[TeV]}
    &
  \Lambda_{\text{NP}}^{\textrm{SFV}} \, \mathrm{[TeV]}
      &  
  \Lambda_{\text{NP}}^{\textrm{MFV}} \, \mathrm{[TeV]}
    \\ \hline
(Q_1^\dagger \bar{\sigma}^\mu Q_2)^2
&
1.5 \times 10^{4} \textrm{\tiny{(Im)}}
&
262.7 \, |\kappa_d^2 - \kappa_s^2|
&
5.1
 \\
(Q_1 \bar{d}_3)(Q_3^\dagger \bar{d}_1^\dagger)
&
2.1 \times 10^{3} \textrm{\tiny{(Abs)}}
&
19.3
 \,
\sqrt{|\kappa_d \kappa_b|}
&
-
\\
(Q_1 \bar{d}_2)(Q_2^\dagger \bar{d}_1^\dagger)
&
2.4 \times 10^{5 } \textrm{\tiny{(Im)}}
&
72.7
 \,
\sqrt{|\kappa_d \kappa_s|}
&
-
    \\
\,
2 e H 
\sigma^{\mu \nu}
Q_2
 \,
  \bar{d}_3 
 \,
 F_{\mu \nu} 
 &
276.3 \textrm{\tiny{(Re)}}
 &
54.3 \, \sqrt{|\kappa_b|}
&
7.0
      \\
\,
2e H 
\sigma^{\mu \nu}
Q_3
 \,
  \bar{d}_2
 \,
 F_{\mu \nu} 
 &
276.3 \textrm{\tiny{(Re)}}
 &
54.3 \, \sqrt{|\kappa_s|}
 &
7.0
      \\
\,
2 eH 
\sigma^{\mu \nu}
Q_3
 \,
  \bar{d}_1
 \,
 F_{\mu \nu} 
 &
140.5 \textrm{\tiny{(Abs)}}
 &
13.2 \, \sqrt{|\kappa_d|}
 &
7.0
\\
\end{array}
$
\end{center}
\caption{
$95\%$ CL bounds on the new physics scale $\Lambda_{\textrm{NP}}$,
for anarchic, SFV and MFV operator coefficients. Bounds are taken from \cite{Bona:2006sa,Bona:2007vi,Crivellin:2011ba,Altmannshofer:2014rta}.
The subscripts on the anarchic operator limits indicates that the limit is on the real, imaginary or absolute value of the operator coefficient. 
For SFV and MFV the phase in the operator coefficient is fixed by the CKM matrix. 
}
\label{tab:operatorbound2}
\end{table}

\section{SFV UV completion}
\label{eq:UVcompletion}

While the SFV Ansatz clearly demonstrates that one can have flavorful new physics at low scales, it is still important to address the assumptions it requires.  
To do so we present one example UV completion for the up-type SFV Ansatz.
A UV completion for the down-type SFV Ansatz can be trivially obtained with the appropriate up-down replacements.
To start, we add to the SM vector-like right handed up-type quarks $U_A,\bar{U}_A$, $A=1..3$,
where $\bar{U}$ has the same gauge quantum numbers as the SM quark $\bar{u}$,
and scalar gauge singlets $S_{iA}$.
In table \ref{tab:charges} we specify the charges of the new quarks and singlets under the different SM and vector-like quark flavor groups.
We introduce interactions between the singlets, vector-like quarks and up-type right-handed  SM quarks.
Our Lagrangian is
\begin{eqnarray}
\nonumber
\mathcal{L}
&\supset&
   M_{\tiny{AB}} U_A \bar{U}_B
   +
   \xi S_{iA} \bar{u}_i U_A 
    \\
    &-&
    \big[
    {\eta}^u_{ij} ~ Q_i H \bar{u}_j
-  {\eta}^{d}_{ij} Q_i H^c \bar{d}_j 
+\textrm{h.c.}
\big]
+
\mathcal{L}_{\textrm{BSM}}
 \label{eq:actionHiggsbasis2}
 \end{eqnarray}
where we omit canonical kinetic terms for all fields and other SM interactions without quarks that are not relevant for our discussion. 
Additional renormalizable interactions to the ones appearing explicitely in Eq. \eqref{eq:actionHiggsbasis2} coupling the vector-like quarks to SM fields may be forbidden by the discrete $\mathbb{Z}_2$ symmetry in Table \ref{tab:charges}.
Without loss of generality, 
we work in a basis where the vector-like quark mass matrix is diagonal $M_{AB}=\delta_{AB} M_A$.
$\mathcal{L}_{\textrm{BSM}}$ represents any other interactions involving arbitrary new physics fields and SM fields,
with the only constraint that no additional spurions or fields transforming under the SM flavor group factor $U(3)_{\bar{u}}$ appear at the renormalizable level.

\begin{table} [ht!]
\begin{center}
$
\begin{array}{c|ccccc}
&
U(3)_U
&
U(3)_{\bar{U}}
&
U(3)_{\bar{u}}
&
U(1)_B
&
\mathbb{Z}_2
 \\
\hline
U
&
3
&
&
&
1/3
&
-1
\\
\bar{U}
&
&
3
&
&
-1/3
&
-1
\\
S
&
\bar{3}
&
&
\bar{3}
&
&
-1
\end{array}
$
\end{center}
\caption{Charge assignments for the vector-like quarks and gauge singlet. SM fields are neutral under the $\mathbb{Z}_2$ symmetry.}
\label{tab:charges}
\end{table}

Next, 
we impose that CP and SM quark family numbers $U(1)_f^3$ are good symmetries in the UV.
In this case, 
there exists a flavor basis in which all spurions transforming under the SM flavor group,
including the SM Yukawa interactions, are real-diagonal $3\times 3$ matrices or tensor products of such matrices. 
In what follows we commit to this real-diagonal flavor basis.
The Yukawa spurions  remain real-diagonal under renormalization from the UV,
protected by the $U(1)_f^3\times \textrm{CP}$ symmetries.
Finally, 
we break $U(1)_f^3\times \textrm{CP}$ softly
only via a VEV for the singlets. 
Note that in this theory the strong-CP problem is solved via the Nelson-Barr mechanism since CP violation is introduced only via mixing with vector-like quarks \cite{Nelson:1983zb,Barr:1984qx,Barr:1984fh,Bento:1991ez}.

\begin{figure} [h!]
\begin{center}
\begin{tikzpicture}[line width=1.5 pt, scale=1.4]
  \draw[fermionbar]   (0, 0.0)-- (-0.8, 0.5);
  \draw[scalarnoarrow]       (-0.8,-0.5)--(0, 0.0);
  \draw[fermionbar]        (0, 0.0)--(1.4, 0.0);
  \draw[fermionbar]    (2.2, 0.5)--( 1.4, 0.0);
  \draw[scalarnoarrow]       (2.2,-0.5)--(1.4, 0.0);
  \node at (0.75, 0.4) {$U_A$};
  \node at (-1.0, 0.6) {$\bar{u}_j$};
  \node at (-1.0,-0.6) {$S_{jA}$};
  \node at ( 2.5, 0.6) {$\bar{u}_i$};
  \node at ( 2.5,-0.6) {$S^{*}_{iA}$};
\end{tikzpicture}
\end{center}
\caption{Tree level diagram leading to the wave-function renormalization operator \eqref{eq:WF}.}
\label{fig:treediag}
\end{figure}
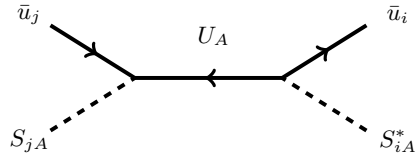 
To understand the effect of the VEV in the infrared,
we treat the singlet condensate as a flavor breaking spurion, 
and we integrate out the vector-like quarks. 
The singlet condensates cannot be $\ll M_{A}$, 
otherwise in the effective theory flavor-changing processes would be much suppressed and the CKM matrix would be close to the identity. 
This motivates defining an operator's effective dimension to be 
\begin{equation}
n_{ED}=4+n_{M^2}-n_{S^2}
\end{equation}
where $n_{M^2}$ and $n_{S^2}$ count powers of vector-like masses and singlet condensate insertions in the operator's coefficient.
The leading effects in the infrared are obtained by working up to effective-dimension four,
higher effective-dimension operators have coefficients suppressed by vector-like quark masses $M_A$ and are dropped in what follows. 
\footnote{Such operators lead to up-type FCNCs so very conservatively they may only be dropped if $M_A > 10^3$ TeV \cite{Bona:2007vi}. 
In our UV completion,
however,
these operators come from diagram \eqref{fig:treediag} at higher order in the momentum expansion.
Using equations of motion, they lead only to effective-dimension six $\Delta F=1$ four fermion operators suppressed by up-type Yukawas,
so in practice the scale of the vector-like quarks may be $\ll 10^3$ TeV. }
In the spurion limit and at tree level, 
the only contributing diagram to the low energy theory is given in Fig. \ref{fig:treediag} (together with diagrams related by gauge invariance). 
At effective-dimension four it leads only to wave-function renormalization of the right-handed up quarks,
so the corresponding low energy theory is given by Eq. \eqref{eq:SFVstructure},
where the up-type quark wave-function renormalization matrix is \footnote{Expression \eqref{eq:WF} is tree-level exact: higher order terms in the effective theory expansion generate short-distance operators but do not lead to corrections to the wave-function matrix.}
\begin{equation}
 Z^u_{ij}=\delta_{ij}+\frac{\xi^* \xi}{M_A^* M_A} S_{iA}^* S_{jA}
 \quad.
 \label{eq:WF}
\end{equation}
$Z^u$ is not diagonal in quark flavor space, and is the only source of breaking of the quark family number and CP groups in the tree level effective-dimension four theory. 
We conclude that this theory is a UV completion leading to up-type SFV at the boundary scale $\Lambda_{\textrm{BC}} \sim M_A$.

With this UV completion we do not attempt to  provide an explanation for the hierarchies of the SM quark masses, mixing angles and CP phase,
which would instead require a full theory of flavor that is beyond the scope of this work. 
This situation is not any worse than in MFV or any other flavorful Ansatz for BSM physics,
which also require a dynamical explanation of the background values of the flavor spurions.
Nevertheless we have verified that the hierarchies in the up-quark masses and CKM elements may be numerically obtained by a combination of hierarchies between the singlet condensates and vector-like quark masses, 
and of the hierarchies of the diagonal Yukawas $\eta^u$ in Eq. \eqref{eq:actionHiggsbasis2}.
The down-type quark masses are independently due to hierarchies in the diagonal down-type Yukawas   $\eta^d$.
We refer the reader to \cite{Nardi:2011st,Espinosa:2012uu,Fong:2013dnk} for an exploration on the dynamical generation of flavored vacua
and to \cite{Davidson:2007si} for more examples of theories generating flavor hierarchies through wave-function renormalization.

\section{Discussion}
\label{sec:discussion}
AFV and SFV provide a promising generalization of the usually flavor universal Ansatz of MFV for BSM physics at the EF.  
Both can be studied from the EFT perspective, 
but the SFV Ansatz is particularly well motivated from UV and IR considerations.
SFV new physics may have large couplings to any of the quark generations, 
leading to novel collider and flavor phenomenology. 
This especially motivates novel collider techniques such as light-quark taggers \cite{Aaboud:2018fhh,Sirunyan:2017ezt,Aaij:2015yqa} at the LHC and beyond.
The SFV Ansatz may be applied to a variety of simplified models, 
as theories with extra Higgses, vector-like quarks or Z-prime bosons \cite{Davidson:2005cw,Grinstein:2010ve,Fox:2011qd,Carone:2013uh}.
A supersymmetric generalization of the SFV Ansatz can also be obtained by promoting the SM fields, vector-like and gauge singlet fields in this work to superfields.
Another interesting future direction is to embed SFV in a complete theory of flavor,
and study how hierarchies different from the SM flavor hierarchies can be realized for different sets of flavor spurions.

\section{Acknowledgements}
We would like to thank Luca Di-Luzio, Howard Haber, Duccio Pappadopulo and Gilad Perez for useful discussions.
The work of DE, SH and PM was supported in part by the National Science Foundation grant PHY-1620628.  
DE and PM thank the Galileo Galilei Institute for Theoretical Physics for the hospitality and the INFN for partial support during the completion of this work, as well as support by a grant from the Simons Foundation (341344, LA).
PM would like to thank the Center for Theoretical Physics at Columbia University for its hospitality during completion of part of this work.
SH is supported by the Department of Energy SCGSR program, administered by the Oak Ridge Institute for Science and Education which is managed by ORAU under contract number DE-SC0014664.

\bibliographystyle{unsrt}
\bibliography{bibliography}

\end{document}